\documentclass[preprint,review,3p,12pt]{elsarticle}

\usepackage{amssymb}
\usepackage{amsmath}
\usepackage{amsthm}

\usepackage{lineno}

\usepackage{graphicx,color}
\usepackage[bookmarks,bookmarksnumbered]{hyperref}
\hypersetup{colorlinks = true,linkcolor = blue,anchorcolor =red,citecolor = blue,filecolor = red,urlcolor = red, pdfauthor=author}
\usepackage{listings}

\definecolor{backcolour}{rgb}{0.95,0.95,0.92}

\begin{document}

\begin{frontmatter}


\title{PennyLane-Lightning MPI: A massively scalable quantum circuit simulator based on distributed computing in CPU clusters}

    \author[a]{Ji-Hoon Kang}
    \author[b]{Hoon Ryu\corref{cor}}

    \cortext[cor]{Corresponding author (\textit{E-mail address:} elec1020@kumoh.ac.kr)}
    \address[a]{Division of National Supercomputing, Korea Institute of Science and Technology Information, Daejeon, 34141, Republic of Korea}
    \address[b]{School of Computer Engineering, Kumoh National Institute of Technology, Gumi, Gyeongsangbuk-do 39177, Republic of Korea}
        
    \begin{abstract}
        Quantum circuit simulations play a critical role in bridging the gap between theoretical quantum algorithms and their practical realization on physical quantum hardware, yet they face computational challenges due to the exponential growth
        of quantum state spaces with increasing qubit size. This work presents PennyLane-Lightning MPI, an MPI-based extension of the PennyLane-Lightning suite, developed to enable scalable quantum circuit simulations through parallelization
        of quantum state vectors and gate operations across distributed-memory systems. The core of this implementation is an index-dependent, gate-specific parallelization strategy, which fully exploits the characteristic of individual gates as well
        as the locality of computation associated with qubit indices in partitioned state vectors. Benchmarking tests with single gates and well-designed quantum circuits show that the present method offers advantages in performance over general
        methods based on unitary matrix operations and exhibits excellent scalability, supporting simulations of up to 41-qubit with hundreds of thousands of parallel processes. Being equipped with a Python plug-in for seamless integration to the
        PennyLane framework, this work contributes to extending the PennyLane ecosystem by enabling high-performance quantum simulations in standard multi-core CPU clusters with no library-specific requirements, providing a back-end resource
        for the cloud-based service framework of quantum computing that is under development in the Republic of Korea.
    \end{abstract}




\end{frontmatter}


    \section{Introduction}\label{sec:intro}
    
    Quantum computing is an emerging paradigm of computation that leverages quantum mechanical principles to address challenges that are difficult to be tackled with classical computing. The fundamental advantages of quantum computing lie
    in the property of quantum bits (qubits), which drives the potential to solve specific tasks much faster than with classical computing, such as factorization~\cite{Shor1994, Shor1997} and searching problems~\cite{Grover1996}. With the aids of
    substantial investments and research efforts, physical quantum computing resources have been realized in various platforms such as superconductors~\cite{Garcia-Perez2020, Karalekas2020} and trapped ions~\cite{Nam2020}. Despite the
    advancement in computing units, the realization of large-scale \& fault-tolerant quantum computers is still hindered mainly by the noise susceptibility of physical qubits, so the practical application of quantum computers to real-world problems
    remains a significant challenge. In consequence, the quantum circuit simulator software, which mimics quantum logic operations in classical high-performance computing (HPC) environments, has become essential for development, verification,
    and evaluation of quantum algorithms as it not only presents a reliable platform to predict the utility of theoretically known fault-tolerant algorithms but also acts as a core component to develop the quantum-classical computing interface dedicated
    to noise-intermediate-scale quantum algorithms for tackling eigenvalue problems~\cite{Peruzzo2014, McClean_2016, vincenzo2021, Cerezo2021, TILLY20221} or combinatorial optimization problems~\cite{farhi2014, farhi2015, Zhou2020,
    Ebadi2022, BLEKOS20241}.
        
    In recent years, various quantum circuit simulators have been developed and released~\cite{smith2016practical, CirqDevelopers2024, qiskit2024, bergholm2022, Jones2019, Guerreschi2020, Amy_2020, Qutip2023}. The most widely adopted
    tools among them are Python-based frameworks such as Forest~\cite{smith2016practical}, Cirq~\cite{CirqDevelopers2024}, Qiskit~\cite{qiskit2024}, and Pennylane~\cite{bergholm2022}, some of which are used as a software development kit
    (SDK) to program quantum circuits with commercial processors. To further enhance computing performance of circuit simulations, several tools are written in modern C++ and optimized to HPC environments, including QuEST~\cite{Jones2019},
    Intel QS~\cite{Guerreschi2020, Smelyanskiy2016}, PennyLane-Lightning~\cite{asadi2024}, Quantum++~\cite{Gheorghiu2018}, and Qulacs~\cite{Suzuki2021,tabuchi2023}. Nevertheless, quantum circuit simulations still remain constrained by the
    exponential growth of the quantum state space with respect to the qubit size that exponentially increases the memory consumption: for instance, simulations of 30-qubit circuits require at least about 16 GB of memory to store a single state vector,
    whereas a 40-qubit system consumes approximately 16 TB of memory. So, partitioning of the state vector across distributed-memory systems becomes essential to address memory issues for simulations of large-scale circuits, requiring parallel
    processing of associated workloads with Message Passing Interface (MPI). 

    Being motivated by statements in previous paragraphs, this work focuses on increasing the scale of quantum circuit simulations in HPC resources by incorporating MPI-based parallelization to the PennyLane-Lightning code~\cite{asadi2024}.
    Being named as ``PennyLane-Lightning MPI'', the proposed tool distributes components of state vectors to different MPI processes and conducts quantum logic operations in parallel with an index-dependent \& gate-specific strategy that exploits
    the data locality of qubits associated with gate operations. This implementation complements the recently published PennyLane-Lightning code~\cite{asadi2024}, which also supports MPI-based parallelization but only works with general purpose
    graphical processing units (GPGPU) since each MPI process conducts circuit simulations with cuQuantum SDK~\cite{bayraktar2023} by directly offloading workloads of gate operations to GPGPU devices. Our tailored approach targets traditional
    multi-core CPU clusters without relying on third-party libraries, and leverages gate-specific optimizations to reduce communication overheads and arithmetic operations. To validate scalability and efficiency of our package for large-scale simulations,
    we demonstrate massive parallelism by conducting quantum circuit simulations with up to hundreds of thousands of MPI processes on the national supercomputer of Korea (NURION system)~\cite{nurion}. We also develop a Python plug-in for
    seamless integration of our code into the existing PennyLane framework~\cite{bergholm2022}, to promote accessibility and usability of the tool for researchers. As an extended version of the well-known PennyLane-Lightning package, our code
    will pave the way to large-scale quantum circuit simulations in distributed-memory computing resources that are not subject to GPGPU environments.

    \section{Implementations}\label{sec:implementation}
    
\begin{figure}[p]
        \centering
        \includegraphics[width=\columnwidth]{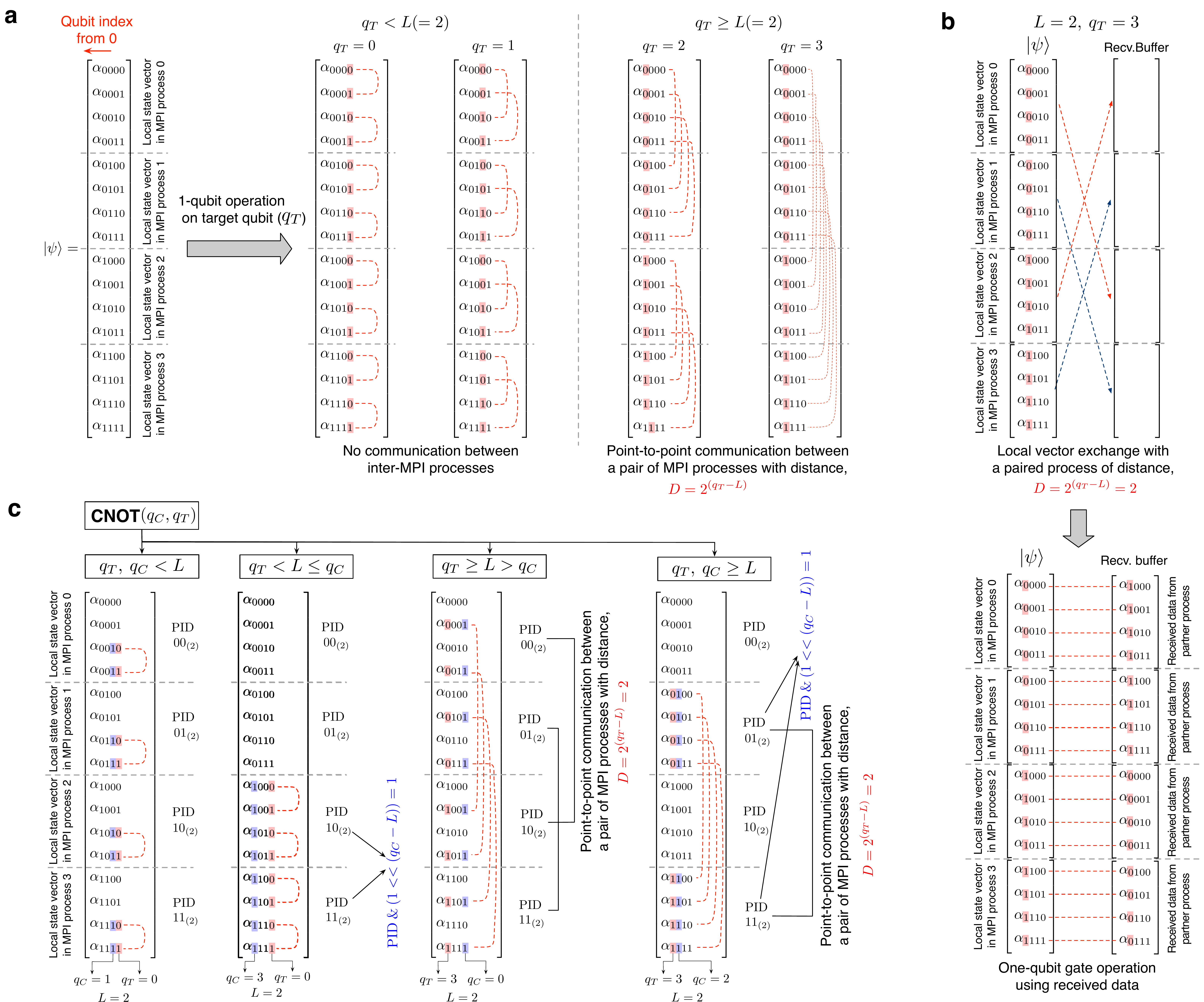}
        \caption{The parallelization scheme for a state vector and gate operations.
        (a) The left panel illustrates the case that a 4-qubit state vector is decomposed across 4 MPI processes. Here, each process stores a $L$-qubit local state vector ($L$ = 2 in this case), and any gate operations to the first $L$ qubits can be simulated in
        parallel with no communication overhead. The right panel shows the case that the target qubit index ($q_T$) of gate operations is greater than or equal to $L$, where simulations of gate operations involve communications between two MPI processes
        at a rank distance of $D = 2^{(q_T - L)}$. (b) If $q_T \ge L$, the paired MPI processes exchange their local vectors and conduct gate operations with both vectors. (c) The index-dependent execution pattern of a CNOT logic is illustrated. Here, $q_T$
        determines whether the parallel operation needs communications as discussed in (a) and (b). The role of the control qubit index ($q_C$) is to determine MPI ranks (PIDs) that participate in conduction of a Pauli-X operation, as shown with blue texts. 
        }
        \label{fig:01}
    \end{figure}

    The PennyLane-Lightning MPI code is designed to exploit the distributed-memory parallelism for scalable simulations of quantum circuits in multi-core CPU clusters. The first issue to be addressed for distributed computing of circuit simulations
    is the strategy to partition the memory-intensive state vector across multiple MPI processes, and here we implement a straightforward scheme where a full $N$-qubit state vector comprising $2^N$ complex amplitudes is evenly distributed among
    $2^p$ processes as done by existing MPI-based simulators~\cite{Jones2019, Guerreschi2020, tabuchi2023, larose2018}. The number of employed MPI processes therefore should always be a power of two, and each MPI process holds a local
    state vector of $2^{N-p}$ amplitudes. This decomposition scheme is well illustrated in the left panel of Figure~\ref{fig:01}(a), which shows the case of a 4-qubit state partitioned with 4 MPI processes (note that qubit index begins with zero and
    increases from left to right). For gate operations, we let all the MPI processes have the corresponding logic matrix since logic matrices representing quantum gates are usually small in size (up to 8$\times$8) so they do not significantly contribute to
    the memory consumption. Accordingly, each MPI process conducts the same gate operation with its own partitioned vector, and we only take nonzero elements of logic matrices to reduce arithmetic operations for gate operations.
    
    In general, simulations of quantum logic operations with distributed computing must involve MPI communications, but the pattern of communications can depend on the target qubit index (indices) against which a gate operation is conducted.
    Let's say the size of each partitioned vector is $2^L$ $(L = N-p)$. Then, any gate operations conducted to the first $L$ qubits of a N-qubit state vector can be simulated in parallel but with no communication between (among) different MPI processes,
    as described in the left panel of Figure~\ref{fig:01}(a) for the case that the target qubit index ($q_T$) is smaller than $L$. In contrast, operations on the remaining $N - L$ qubits ($i.e.,$ $L$ $\leq$ $q_T$ $<$ $N$) involve MPI communications
    as shown in the right panel of Figure~\ref{fig:01}(a) for the case of $q_T \ge L$. Accordingly, we define qubits of indices $0 \leq q < L$ as $local$ $qubits$, and those of indices $L \leq q < N$ as $non$-$local$ qubits. Using this definition, we
    determine the communication pattern and conduct MPI communications only when the target qubit is non-local. When the target qubit is non-local, each MPI process exchanges its local state vector with its paired process at a rank distance
    of $D=2^{(q_T-L)}$, and executes gate operations using its own and received vector, as illustrated in Figure~\ref{fig:01}(b). The pattern of communications also depends on the characteristic of a quantum logic gate. For example, though both
    Pauli-X and Pauli-Z are single-qubit logic, the Pauli-X gate needs MPI communications when the target qubit is non-local while the Pauli-Z one is completely free from communications since it just changes amplitudes of a state vector at indices
    where the target qubit is 1. Figure~\ref{fig:01}(c) shows the case of a controlled-X (CNOT) gate that has four communication patterns depending on indices of a control and a target qubit. If the target qubit is local (1st and 2nd case), no communication
    is necessary regardless of the control qubit index; in such case, the operation can be completed in each MPI process (if the control qubit is local) or by conducting a Pauli-X operation in MPI processes determined by the control  qubit (if the control
    qubit is non-local). If the target qubit is non-local (3rd and 4th case), simulations always involve communication, through which a Pauli-X logic is conducted in MPI ranks determined by the control qubit index. To reduce the communication cost for
    simulations as much as possible, we avoid unnecessary communications by incorporating the gate-specific characteristic into the index-dependent strategy for determination of the communication pattern.
    
    It is worth noting that QuEST~\cite{Jones2019} and Intel QS~\cite{Guerreschi2020} have proposed a memory-efficient strategy in sacrifice of communication overhead, $i.e.,$ MPI processes exchange only parts of local state vectors through
    a single communication and conduct multiple communications to complete simulations in parallel. But, here we do not consider such memory optimization and complete simulations with a single round of communication if needed. This simple
    choice of MPI implementation is based on the fact that many gate operations can be conducted in parallel with no communications as discussed in the previous paragraph. In addition, our implementation includes the Python binding interface
    dedicated to the routines developed for gate operations and measurement processes, so, through the use of the \texttt{lightning\_mpi.qubit} backend that extends the existing \texttt{lightning.qubit} one to MPI environments, users can build parallel
    workloads with minimal changes of their original sequential Python codes. As shown in Listing~1, the implementation integrates with \texttt{mpi4py}~\cite{dalcin2021}, providing a user-friendly interface for MPI initialization and rank management.
    To evaluate the parallel performance of the proposed implementation, we conducted benchmark tests on the NURION supercomputer operated by the Korea Institute of Science and Technology Information (KISTI) \cite{nurion}. The NURION system
    consists of 8,305 Cray CS500 computing nodes, where each node is equipped with an Intel Xeon Phi 7250 processor (68 cores) and 96 GB of DDR4 main memory along with 16 GB of high-bandwidth on-chip memory (a total main memory of
    $\sim$780 TB). The entire system is interconnected via the Intel Omni-Path Architecture, enabling efficient data communication across the system.

    \section{Results and discussion}\label{sec:Results}

    \subsection{Performance analysis: Individual gate operations}\label{sec:gate}
     
\begin{figure}[p]
        \centering
        \includegraphics[width=\columnwidth]{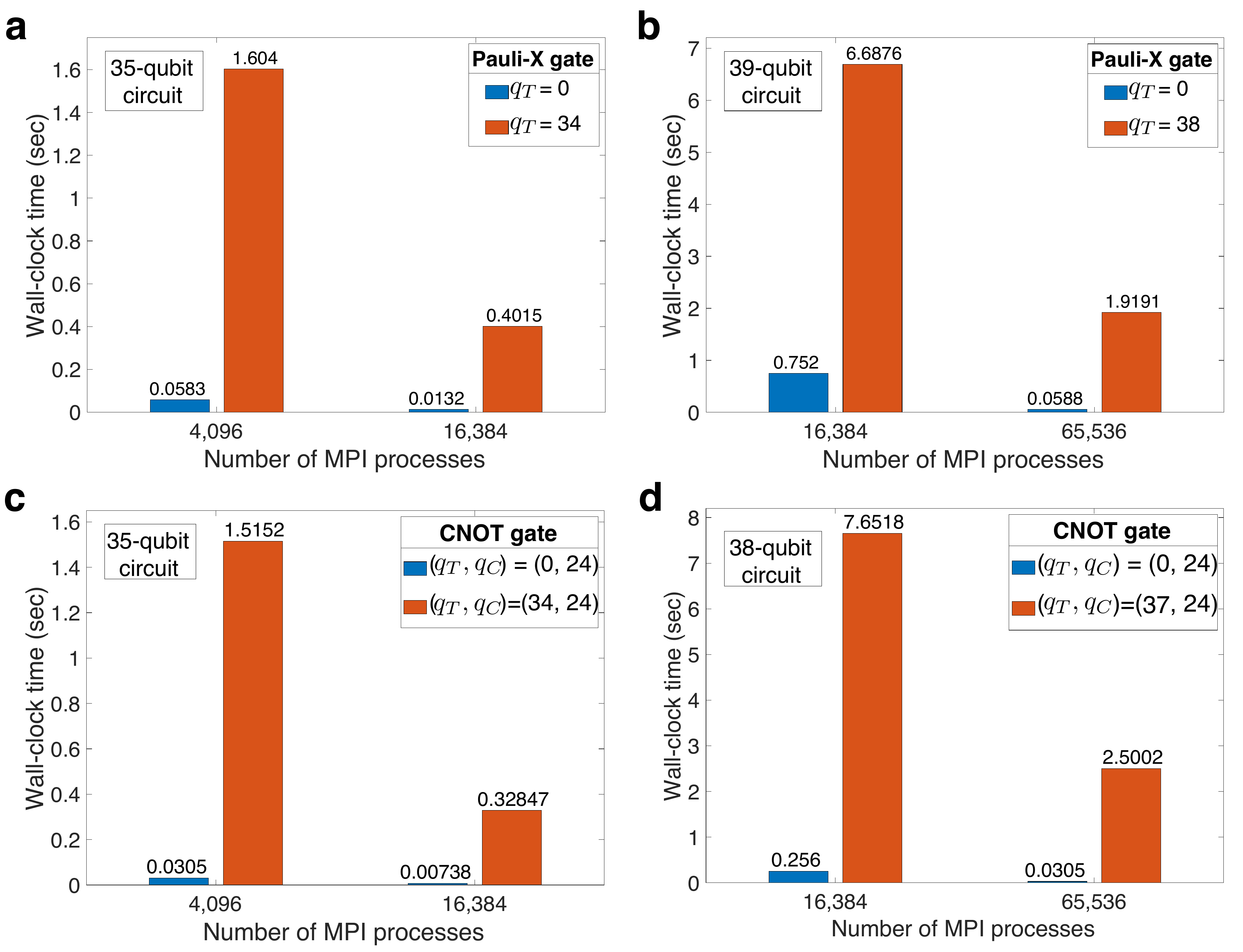}
        \caption{Performance of local and non-local operations.
        (a) \& (b) show the wall-clock time of a single Pauli-X operation that is conducted in a 35-qubit \& a 39-qubit circuit, respectively. When the gate operation is conducted to the first qubit ($q_T$ = 0), all the operations become local and are completed
        much faster than the wort case where $q_T$ is the maximum value ($q_T = 34$ in (a), $q_T = 38$ in (b)) so all the operations become non-local involving MPI communications. (c) \& (d) show the wall-clock time of a CNOT operation that is conducted
        in a 35-qubit \& a 38-qubit circuit, respectively. In both cases, operations involving the maximum communication distance ($q_T = 34$ in (c), $q_T = 37$ in (d)) takes remarkably longer times than fully local ones ($q_T = 0$).
        }
        \label{fig:02}
    \end{figure}

    We first elaborately analyze the performance of PennyLane-Lightning MPI with representative single-gate operations, particularly to rigorously evaluate the effectiveness of our index-dependent parallelization strategy that is discussed in the
    section \ref{sec:implementation}. For individual quantum gates subjected to benchmark tests, we consider a single-qubit Pauli-X gate and a two-qubit CNOT gate since they well represent the family of single-qubit and two-qubit universal gates
    that must involve MPI communications when employed by simulations of large-scale quantum circuits. In Figure \ref{fig:02}(a), we present the wall-clock time that is measured for the single-qubit Pauli-X operation conducted to various target
    qubits in a 35-qubit circuit. Results here show a significant variation in execution time depending on the target qubit index such that, when $q_T$ = 0, the operation becomes fully local and $embarrassingly$ $parallel$. In contrast, the operation
    with the highest $q_T$ ($=$ 34) conducts communications between all the pairs of MPI ranks that are separated by half of the total size of processes. Notably, the Pauli-X operation with $q_T$ = 34 turns out to be $\sim$30 times slower than
    that with $q_T$ = 0 regardless of the size of MPI processes, clearly indicating the remarkable cost of MPI communications that should be paid for non-local operations of even a single-qubit gate. Similar messages can be drawn by Figure
    \ref{fig:02}(b), which shows the wall-clock time measured for 39-qubit circuit simulations. Results of a single CNOT operation are presented in Figures \ref{fig:02}(c) and \ref{fig:02}(d), which show the wall-clock time taken to complete simulations
    of a 35- and a 38-qubit circuit, respectively. Here in all the cases we use a non-local control qubit fixed with an index of 24 to make the communication pattern solely depend on the target qubit index. Results show that the operation with $q_T$
    = 0 (most local) is up to 80 times faster than the one with highest $q_T$'s (most non-local), confirming that our strategy remarkably enhances the performance of circuit simulations by avoiding unnecessary communications. 
    
\begin{figure}[p]
        \centering
        \includegraphics[width=\columnwidth]{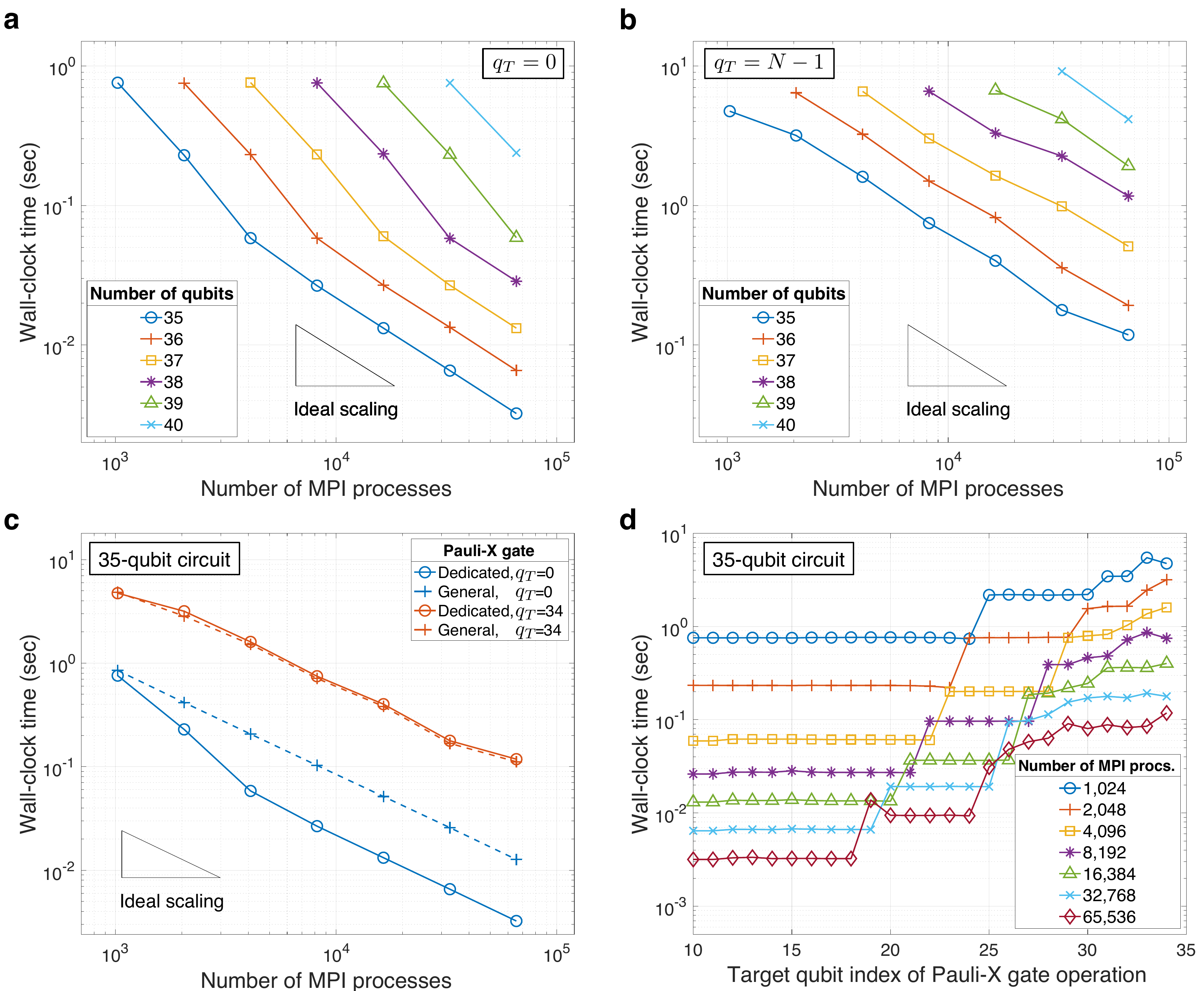}
        \caption{Performance of a single-qubit operation in large circuits.
        (a) \& (b) show the strong scalability of our code when a single Pauli-X gate is applied to the first \& the last qubit in 35-to-40 qubit circuits, respectively. In both cases, the code demonstrates fairly nice scaling behaviors, but the performance obtained
        with the first qubit ($q_T$ = 0) is better in terms of parallel efficiency and computing time than the case for the last qubit ($q_T = N-1$, $N$ = the circuit size). (c) Our index-dependent parallelization strategy (``Dedicated'') notably reduces the wall-clock
        time for local operations ($q_T = 0$) against the results driven by the general approach (``General'') that conducts gate operations with matrix-vector multiplications. When $q_T = 34$, the performance gain vanishes as all the operations in our method
        become non-local, losing the strength in communication overheads. (d) The wall-clock times of simulations conducted in a 35-qubit circuit (64 MPI processes per node) are shown as a function of $q_T$. Two distinct jumps are observed as $q_T$ increases:
        the $1^{st}$ and the $2^{nd}$ one are due to the overhead of intra- and inter-node communications, respectively.
        }
    	\label{fig:03}
    \end{figure}

    The scaling behavior of single-gate operations is also examined to evaluate the parallel efficiency of our code. Figure~\ref{fig:03}(a) and \ref{fig:03}(b) show the scaling curves of 35-to-40-qubit circuits, where a single Pauli-X gate is applied to
    the first and last qubit, respectively. In general, both cases show fairly nice scaling behaviors, but, when the operation is fully local (Figure~\ref{fig:03}(a)), the super-linear graph is observed regardless of circuit sizes due to the improved memory
    access time driven by the reduced memory usage per MPI process. This behavior is much less remarkable in non-local operations due to the communication overhead, as can be confirmed from Figure~\ref{fig:03}(b). The strength of our gate-
    \& index-specific parallel simulations is also huge when compared to the general approach that represents the Pauli-X (any single-qubit) gate with a $2\times2$ unitary matrix. For the fully local operation ($q_T = 0$), our implementation remarkably
    outperforms the general approach, and the advantage in performance becomes more pronounced as more MPI processes are involved as Figure~\ref{fig:03}(c) shows. The advantage of our approach reduces as $q_T$ increases, and eventually
    vanishes at the maximum $q_T$ where simulations are dominated by communications. Figure~\ref{fig:03}(d) shows the execution time of a Pauli-X operation in a 35-qubit circuit as a function of $q_T$ and the number of employed MPI processes
    where 64 (= $2^6$) processes are created per computing node. With no exception, a distinct jump in execution time is observed at $q_T = N - p$ ($N = 35$, $2^p$ = the number of total MPI processes) where intra-node communications start to
    happen. The 2nd increase occurs at $q_T = N - (p - 6)$ where simulations start to involve inter-node communications.

\begin{figure}[p]
        \centering
        \includegraphics[width=\columnwidth]{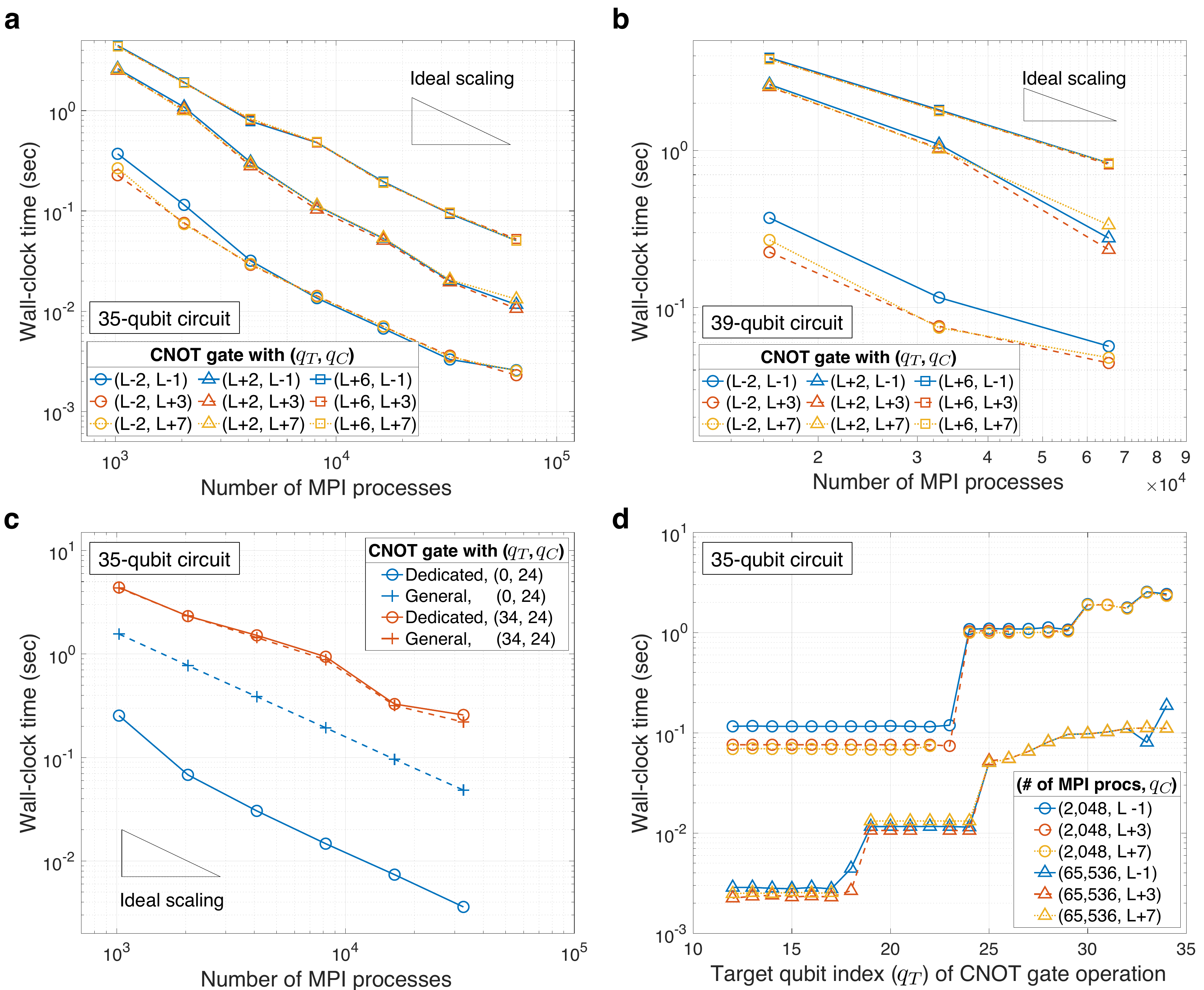}
        \caption{Performance of a two-qubit CNOT operation in large circuits. (a) \& (b) show the strong scalability of our code when a CNOT operation is conducted in a 35- \& a 39-qubit circuit, respectively ($L$ = the size of local qubits). In both cases, the target
        qubit ($q_T$) significantly affects the performance of simulations while the control qubit ($q_C$) does not. The scaling behaviors are also quite nice, being close to the ideal one. (c) When the operations are fully local with $(q_T, q_C) = (0, 24)$, our
        parallelization strategy (``Dedicated'') remarkably outperforms the general one that describes the CNOT gate with a 4$\times$4 matrix. This advantage in performance reduces as $q_T$ increases, and disappears when $(q_T, q_C) = (34, 24)$ where
        inter-node communications dominate the execution time. (d) The wall-clock times of simulations, which are conducted in a 35-qubit circuit with 64 MPI processes per node, are shown as a function of $q_T$, where two noticeable jumps in time are observed
        due to the initiation of intra- and inter-node communications.
        }\label{fig:04}
\end{figure}

    The scalability of a two-qubit CNOT gate is shown in Figure~\ref{fig:04}(a) (in a 35-qubit circuit) and \ref{fig:04}(b) (in a 39-qubit circuit), where the control qubit index ($q_C$) is parameterized as well. Similarly to the case of Pauli-X operations,
    all the scalabilities here are quite nice, being close to the ideal value. The execution time also hugely depends on $q_T$ like the single-qubit Pauli-X case, but its sensitivity to $q_C$ becomes much weaker since a non-local control qubit does
    not trigger communications unlike what the target qubit does; it just determines which MPI processes need to conduct the Pauli-X operation. The advantage of our implementation in terms of computing time is also investigated against the general
    approach that handles the CNOT logic with a 4$\times$4 unitary matrix. The pattern of results in Figure~\ref{fig:04}(c) is quite similar to that of the single-qubit Pauli-X case (Figure~\ref{fig:03}(c)), and notable increase in computing speed is
    observed for local operations. However, the performance gap between our approach and the general one becomes more pronounced in this case, particularly when the control qubit is non-local while the target qubit remains local ($(q_T, q_C)
    = (0, 24)$ in Figure~\ref{fig:04}(c)). In this condition, our index-dependent strategy avoids unnecessary communications as the control qubit only determines the process ranks that conducts NOT operation. In contrast, the matrix-based approach
    redundantly performs MPI communications despite they have no effects on the computation. Finally, the wall-clock time required to complete a CNOT operation in a 35-qubit circuit is presented in Figure~\ref{fig:04}(d) as a function of $q_T$,
    $q_C$, and the number of participating MPI processes ($2^6$ processes per node). Like what is observed from Figure~\ref{fig:03}(d), here we find two points of $q_T$ where the wall-clock time sharply increases due to the overhead of intra-
    \& inter-node communications.

    \subsection{Performance analysis: Quantum circuit operations}\label{sec:circuit}

\begin{figure}[p]
        \centering
        \includegraphics[width=\columnwidth]{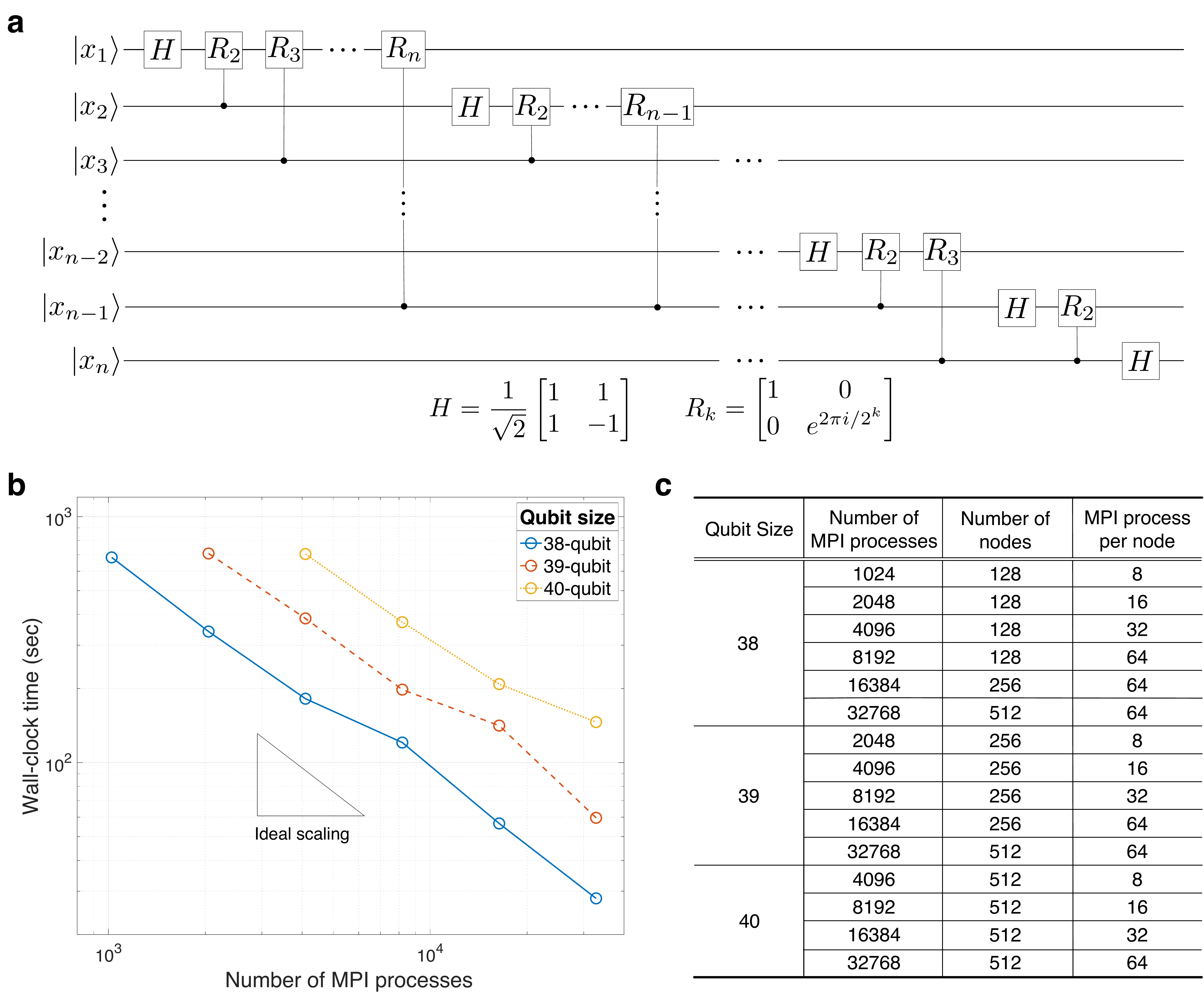}
        \caption{Performance of algorithmic operations - the quantum Fourier transform (QFT) circuit. (a) The QFT circuit is composed of Hadamard ($H$) gates and controlled phase gates, where $R_k$ is the single-qubit rotation about the computational (Z) axis
        by $\pi$/2$^{k}$ radian. (b) The wall-clock times for simulations of 38-to-40 qubit QFT circuits are shown as a function of the number of MPI processes. In general, the observed scaling behaviors are fairly excellent, being quite close to the ideal one. (c) The
        table shows the number of computing nodes and MPI processes that are employed to conduct parallel simulations of 38-to-40 qubit QFT circuits.
        }\label{fig:05}
    \end{figure}

    To examine the parallel scalability of our simulator in the algorithmic circuit level, here we conduct benchmark tests with the two well-established tasks: one is the Quantum Fourier Transform (QFT) that belongs to most popularly employed
    fault-tolerant algorithms, and the other is the universal quantum circuit that can represent the most expensive workload and is known to be capable of generating any arbitrary quantum states upon the selection of parameterized angles of
    single-qubit rotations~\cite{Sousa2007}. Figure \ref{fig:05}(a) describes the QFT circuit that sequentially conducts single-qubit Hadamard and two-qubit controlled-Phase operations, and the execution times of 38-to-40-qubit circuits are presented
    in Figure~\ref{fig:05}(b) as a function of the number of MPI processes participating in simulations. Due to the substantial memory footprint of large-scale circuits, we adjust the number of MPI processes per computing node to ensure the memory
    availability, where detailed configurations for benchmark tests are described in Figure~\ref{fig:05}(c). Figure~\ref{fig:05}(b) clearly supports that all the workloads have excellent scaling behaviors regardless of qubit sizes, though minor degradation
    in performance, $i.e.,$ the slope of a scaling curve, is consistently observed when each computing node uses 32-to-64 processes, due to the reduced memory bandwidth per process as memory usage approaches the hardware limitation.

\begin{figure}[p]
        \centering
        \includegraphics[width=\columnwidth]{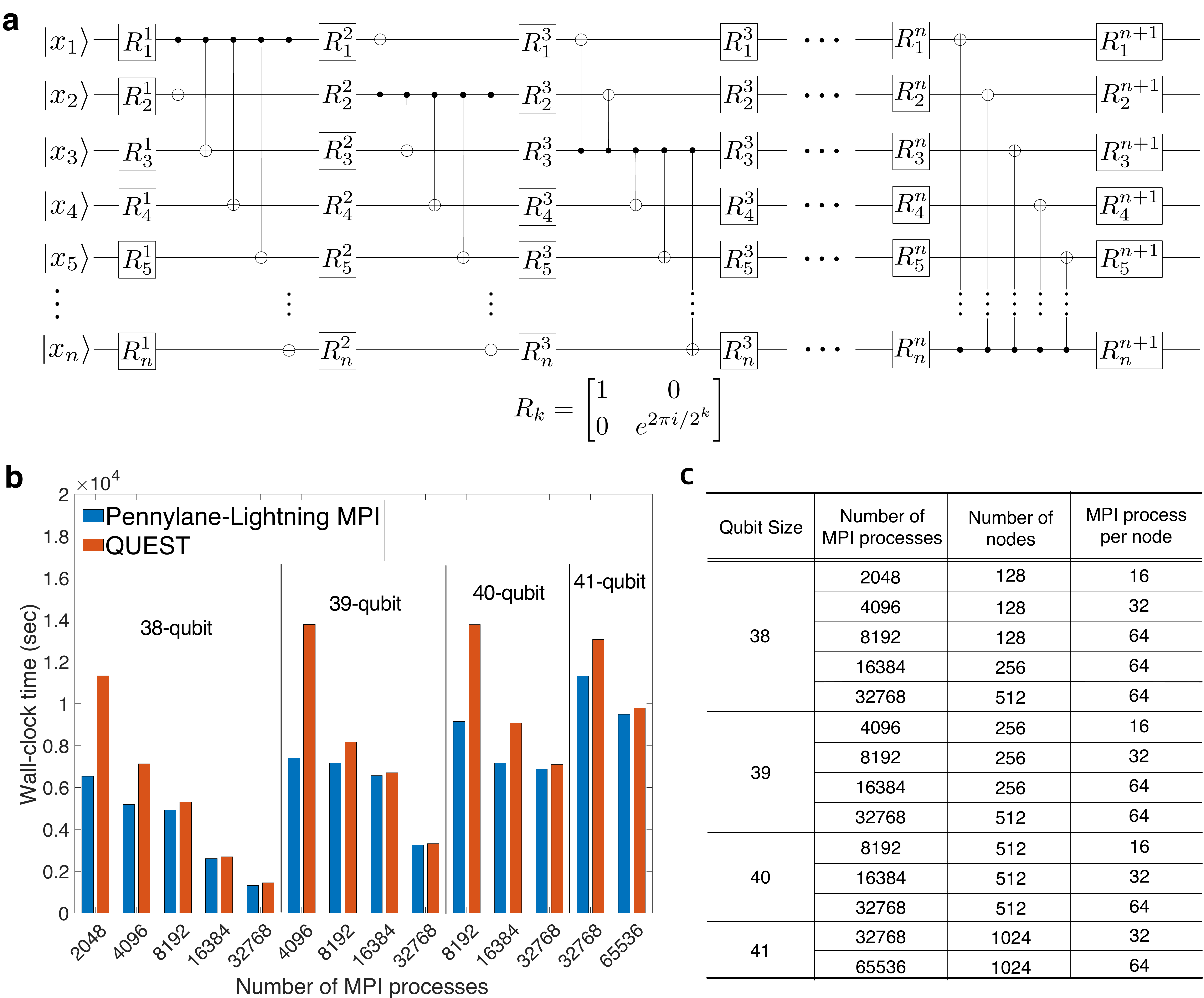}
        \caption{Performance of algorithmic operations - the universal quantum circuit. (a) Being composed of single-qubit gates and CNOT gates, the $N$-qubit universal circuit can generate any $N$-qubit states depending on the parameterized angles of single-qubit
        rotations. Here, we test a specific case of 38-to-41 qubit universal circuits where phase gates $R_k$'s are employed as single-qubit rotations. (b) The wall-clock times of circuit simulations conducted with our code (blue bars) and the QuEST simulator (orange
        bars) are presented as a function of the number of MPI processes. With fairly nice scaling behaviors, our code consistently takes less execution times than QuEST for all the workloads of benchmark tests. (c) The table shows the number of computing nodes
        and MPI processes that are employed to conduct parallel simulations of 38-to-41 qubit universal quantum circuits.
         }\label{fig:06}
    \end{figure}

    Figure~\ref{fig:06}(a) shows the universal quantum circuit that can represent an arbitrary $N$-qubit logic gate, where the logic $R_{k}$ indicate the single-qubit rotation about the computational (Z) axis by $\pi$/2$^{k}$ radian. The $N$-qubit
    universal circuit considered here has a total of 2$N^2$ gates ($N\times$($N-1$) CNOTs and $N\times$($N+1$) $R_{k}$'s) with a depth of $N^2-1$, and we examine the parallel efficiency of our code with 38-to-41 qubit circuits. In Figure~\ref{fig:06}(b),
    we show the results of benchmark tests with those driven by the QuEST quantum simulator~\cite{Jones2019}, where corresponding details of the MPI setup are described in Figure~\ref{fig:06}(c). In terms of the strong scalability, QuEST turns
    out to be a bit better than our code, but this is mainly due that QuEST needs more time to complete simulations particularly when a smaller number of MPI processes are employed. In general, our PennyLane-Lightning MPI code shows better
    performance in computing time than QuEST regardless of the number of employed MPI processes (up to 80\% faster computing time), ensuring the effectiveness of our index-dependent, gate-specific parallelization strategy. Overall, the results
    obtained from benchmark tests so far clearly confirm that PennyLane-Lightning MPI achieves efficient and scalable performance for both single-gate and circuit-level simulations in distributed-memory systems.

    \section{Conclusion}

    We have introduced the PennyLane-Lightning MPI code, an MPI-based extension of the PennyLane-Lightning suite designed for scalable simulations of large quantum circuits in computing environments based on distributed-memory architectures.
    An index-dependent, gate-specific parallelization scheme is devised and implemented to partition full quantum state vectors across MPI processes with reduced computing and communication overhead. Benchmark tests have been rigorously
    conducted with gate-level and circuit-level simulations in the national supercomputer of the Republic of Korea, and excellent scalabilities have been confirmed against up to 41-qubit workloads. Results reveal that the communication overhead,
    which strongly depends on qubit indices on which gate operations are conducted, significantly affects the execution time of large-scale simulations, and our tailored parallelization scheme clearly drives excellent performance by suppressing
    redundant communications particularly when simulations are conducted with a massive number of MPI processes. The advantage of our parallelization scheme in terms of computing performance has been also confirmed through elaborated
    comparisons to the well-known QuEST simulator. Being equipped with a Python plug-in extension and independent of any third-party libraries, our code not only serves as a complementary backend for PennyLane-Lightning in conventional
    multi-core CPU clusters, but also provides a lightweight \& portable solution to harness high-performance computing for quantum circuit simulations.

    \section*{Acknowledgements}
    This work has been carried under the support from the National Research Foundation of Korea (Grant \#: RS-2022-NR068791) and the Korea Institute of Science and Technology Information (KISTI) (Grant \#: K25L2M2C3). Authors acknowledge
    the extensive utilization of the NURION supercomputer in KISTI and the high performance computing resources in the Supercomputing Center of the Kumoh National Institute of Technology (2025).

    \bibliographystyle{elsarticle-num}
    \bibliography{Manuscript_PLL_MPI}

    \newpage
    \begin{lstlisting}[language=Python, caption= An exemplary variational quantum eigensolver code developed with our \texttt{lightning\_mpi.qubit} backend and Python plugin. 
    			     The python-level programming of quantum circuits can be done in the same way as it is done with the \texttt{lightning.qubit} backend.]
import pennylane as qml
from pennylane import numpy as np
from mpi4py import MPI

comm = MPI.COMM_WORLD
rank = comm.Get_rank()

symbols = ["H", "H"]
coordinates = np.array([0.0, 0.0, -0.6614, 0.0, 0.0, 0.6614])
electrons = 2

H, qubits = qml.qchem.molecular_hamiltonian(symbols, coordinates)
dev = qml.device("lightning_mpi.qubit", wires=qubits, mpi=True)
hf = qml.qchem.hf_state(electrons, qubits)

def circuit(param, wires):
    qml.PauliX(wires=0)
    qml.PauliX(wires=1)
    qml.DoubleExcitation(param, wires=[0, 1, 2, 3])

@qml.qnode(dev, interface="autograd")
def cost_fn(param):
    circuit(param, wires=range(qubits))
    return qml.expval(H)

opt = qml.GradientDescentOptimizer(stepsize=0.4)
theta = np.array(0.0, requires_grad=True)
energy = [cost_fn(theta)]
angle = [theta]
max_iterations = 100
conv_tol = 1e-06

for n in range(max_iterations):
    theta, prev_energy = opt.step_and_cost(cost_fn, theta)
    energy.append(cost_fn(theta))
    angle.append(theta)
    conv = np.abs(energy[-1] - prev_energy)
    if conv <= conv_tol:
        break

if rank == 0 : 
    print("\n" f"Obtained ground-state energy = {energy[-1]:.8f} Ha")
    print("\n" f"Optimal circuit parameters = {angle[-1]:.4f}")
    \end{lstlisting}

    \clearpage
    
\end{document}